\documentclass[11pt]{article}
\usepackage{amsmath}
\usepackage{graphicx}
\usepackage{amssymb}
\usepackage{amsfonts}
\usepackage{latexsym}
\renewcommand{\baselinestretch}{1.1}
\def\beq{\begin{eqnarray}}
\def\eeq{\end{eqnarray}}

\def\det{\,\mbox{det}\,}

\def\diag{\,\mbox{diag}\,}

\def\la{\lambda}

\begin{document}
\begin{center}
{\Large\bf BIREFRINGENCE PHENOMENA REVISITED}
\vskip 12mm
\textbf{Dante D. Pereira\footnote{
dante.pereira@cefet-rj.br },
\
Baltazar J. Ribeiro\footnote{
baltazarjonas@nepomuceno.cefetmg.br}
\ and Bruno Gon\c calves\footnote{
bruno.goncalves@ifsudestemg.edu.br}}
\vskip 8mm
$^{1}$Centro Federal de Educa\c c\~ao Tecnol\'ogica Celso Suckow da Fonseca
 \\CEFET-RJ,  27.600-000, Valen\c ca, Rio de Janeiro, Brazil\\
$^{2}$Centro Federal de Educa\c c\~ ao Tecnol\' ogica de Minas Gerais \\ 
CEFET-MG, 37.250-000, Nepomuceno, Minas Gerais, Brazil\\
$^{3}$Instituto Federal de Educa\c c\~ ao, Ci\^ encia e Tecnologia do Sudeste de Minas Gerais
\\ IF Sudeste MG, 36080-001, Juiz de Fora, Minas Gerais, Brazil
\end{center}
\vskip 4mm
\begin{quotation}
\begin{abstract}
The propagation of electromagnetic waves is investigated in the context of the isotropic 
and nonlinear dielectric media at rest in the eikonal limit of the geometrical optics. 
Taking into account the functional dependence $\varepsilon=\varepsilon(E,B)$ and $\mu=\mu(E,B)$ 
for the dielectric coefficients, a set of phenomena related to the birefringence of the 
electromagnetic waves induced by external fields are derived and discussed.  
Our results contemplate the known cases already reported in the literature: 
Kerr, Cotton-Mouton, Jones  and magnetoelectric
effects. Moreover, new effects are presented here 
as well as the perspectives of its experimental confirmations.
\end{abstract}
\end{quotation}
\vskip 4mm
\section{Introduction}
\hspace{0.65cm}Electromagnetic waves in nonlinear media propagate according to Maxwell's equations complemented by
certain phenomenological constitutive relations linking strengths and induced fields \cite{ref1}. Depending on the dielectric
properties of the medium and also on the presence of applied external fields, a variety of optical effects can be
found. One of such an effects which has received significant attention of the scientific community in the last 
years is the birefringence phenomenon (or duble refraction) \cite{ref2, ref3}. This effect takes place when the electromagnetic waves propagate in a medium which exhibits two distinct refractive indexes, in a same wave vector direction. Birefringence was first described by Rasmus Bartholin in 1669, who observed it in calcite (natural birefringence) \cite{ref6}. Moreover, the application of external fields in a medium with nonlinear dielectrics properties, may generate an artificial optical axis which turn the material into optically anisotropic 
(artificial birefringence) \cite{ref5}.

Birefringence induced by external electromagnetic fields in nonlinear media has been reported in the literature since a 
long time ago. Historically, Kerr effect was discovered in 1875 where a relation between optics and electrostatic fields was founded through the observation of the birefringence in glasses and similar observations in liquids \cite{ref7}. Some years later, Cotton and Mouton observed an identical phenomenologically effect for magnetic fields \cite{ref8}. In both cases, birefringence effect is squared with respect to the intensity of the external field applied and the optical axis is parallel to the direction of these fields. The possibility of a combined effect with a simultaneously action of electric and magnetic field has led to the discovery of the Jones effect in 1948 \cite{ref9}. Jones showed that this effect could exist only in uniaxial media and it represents an additional case of the standard birefringence. Since then, Jones formalism has been considered as an useful tool in order to describe the light propagation in media with one or more possible distinct optical effects. Recently, new kinds of magnetoelectric optical effects has been experimentally and analytically investigated \cite{ref11, ref19}. Several applications of the birefringence effects are regarded in the papers \cite{ref12}.

The main point in this work is the search for new effects. In order to present a more general description of the birefringence phenomenon, 
we deal with isotropic and nonlinear dielectric media at rest characterized by the dielectric coefficients $\varepsilon=\varepsilon(E,B)$ and $\mu=\mu(E,B)$ in the eikonal limit of geometrical optics \cite{ref4}. The analysis is restricted to local electrodynamics, where dispersive
effects are neglected. Only monochromatic waves are considered and therefore there are no 
ambiguities with the velocity of the waves.

The paper is organized as follows. In the next section, through the shock waves formalism, we derive the eigenvalues equation related to the 
light propagation in dielectric nonlinear isotropic media. Cumming up, we get the dispersion relations to the possible propagating polarization modes. We derived a birefringence equation which presents three new optical birefringent phenomena. 
Each effect is analyzed separately. Such an equation also contemplates the well known five cases reported 
in the literature: Kerr, Cotton-Mouton, Jones and magnetoelectric effects. In Sec. 4 we discuss the possible experimental perspectives in order to measure these three new effects. In Sec. 5 we draw our conclusions. Aspects about effective geometry and polarization modes are considered in the Appendix. A Minkowskian spacetime described in an adapted Cartesian coordinate system is used throughout this work. The units are such that $c = 1$. The spacetime metric is denoted by $\eta_{\alpha\beta} = \diag(+1,-1,-1,-1)$. All quantities are referred to as measured by the geodetic observer $V^{\alpha}=\delta^{\alpha}{}_{0}$, where $\delta^{\alpha}{}_{\beta}$ denotes the Kronecker tensor.
\section{Nonlinear electrodynamics: the Fresnel generalized equation}

\hspace{0.65cm}Let us begin by mentioning that 
the dynamics of the electromagnetic Maxwell's theory is described,  
in the absence of sources, 
by the set of coupled first order hyperbolic partial 
differential equations \cite{ref2, ref3} 
\begin{eqnarray} 
\partial_{\beta}P^{\alpha\beta}=0\,,  \hspace{1cm}
\eta^{\alpha\beta\gamma\delta}\partial_{\gamma}F_{\alpha\beta}=0 \,,
\label{Maxwell Eqs.}
\end{eqnarray}
where the Levi-Civita tensor $\eta^{\alpha\beta\gamma\delta}$ is defined 
as $\eta^{0123} =+1$ 
and the antisymmetric tensors $F^{\alpha\beta}$ 
and $P^{\alpha\beta}$ represent 
the total electromagnetic field 
\cite{ref2}. Such fields can be expressed 
in terms of the intensities 
$({E}^{\alpha},{B}^{\alpha})$ 
and excitations $({D}^{\alpha},{H}^{\alpha})$ of the electromagnetic 
fields\footnote{Electromagnetic fields 
shall be here represented by 4-vectors space-like  $E^{\alpha}
=(0,\vec{E})$ and $B^{\alpha}=(0,\vec{B})$, where 
internal and external products are defined as 
$\eta_{\alpha\beta}E^{\alpha}B^{\beta}
=-\vec{E}\cdot\vec{B}$ 
and $\eta^{\alpha\beta\gamma\delta}E_{\beta}V_{\gamma}B_{\delta}
=(\vec{E}\times\vec{B})^{\alpha}$, respectively.} in the following way 
\begin{eqnarray} 
F^{\alpha\beta}= V^{\alpha}E^{\beta} - V^{\beta}E^{\alpha} 
- \eta^{\alpha\beta}{}_{\gamma\delta}V^{\gamma}B^{\delta}\,,  
\hspace{0.8cm}
P^{\alpha\beta}= V^{\alpha}D^{\beta} - V^{\beta}D^{\alpha} 
- \eta^{\alpha\beta}{}_{\gamma\delta}V^{\gamma}H^{\delta} \,.
\label{electromagnetic tensors}
\end{eqnarray}  
In the last equation, $V^{\alpha}=\delta^{\alpha}{}_{0}$ represents the 
velocity 4-vector in the framework of an observer at rest in the laboratory, where 
the fields are measured. 
According to the material media approach, the 
dielectric properties of the medium are determined by the matrices 
$\varepsilon^{\alpha}{}_{\beta}({E}^{\gamma},{B}^{\gamma})$ and 
$\mu^{\alpha}{}_{\beta}({E}^{\gamma},{B}^{\gamma})$. These matrices
are related to the intensity fields and to the excitations by the 
phenomenological constitutive relations \cite{ref3} 
\begin{equation}
D^{\alpha}=\varepsilon^{\alpha}{}_{\beta}E^{\beta}\,,
\hspace{0.8cm}
H^{\alpha}
=\mu^{\alpha}{}_{\beta}B^{\beta}\,.
\end{equation}
In the nonlinear isotropic media context, the dielectric matrices 
may be written as \cite{ref13}
\begin{eqnarray}
\varepsilon^{\alpha}{}_{\beta} = \varepsilon h^{\alpha}{}_{\beta} 
= (\varepsilon_{0} 
+ \varepsilon_{ind}) h^{\alpha}{}_{\beta} ,\;\;\;\;
\mu^{\alpha}{}_{\beta} = {\mu}^{-1} h^{\alpha}{}_{\beta} = 
(\mu_{0} + \mu_{ind})^{-1} h^{\alpha}{}_{\beta} \,,
\label{constitutive relations 1}
\end{eqnarray}
where $h^{\alpha}{}_{\beta}:=\delta^{\alpha}{}_{\beta} 
- V^{\alpha}V_{\beta}$ is the projector in the observer's 3-dimensional space $V^{\alpha}$. 
The dielectric coefficients are composed by two parts ($\varepsilon_{0}, \;\mu_{0}$) and 
($\varepsilon_{ind}, \;\mu_{ind}$). 
The first one is a background contribution part 
(inherent to the material) and the second 
one represents an induced contribution\footnote{Such a contribution is related to the external agents applied to the medium.}. 
For the model considered here, the quantities $\varepsilon_{ind}$ and $\mu_{ind}$ 
depend on the external electromagnetic fields 
applied to the material in the following way 
\begin{eqnarray} 
\varepsilon_{ind} (E,B) = \varepsilon_{1}E^{2} + \varepsilon_{2}B^{2} 
+ \varepsilon_{3}(\vec{E}\cdot\vec{B}), \;\;\;\;
\mu_{ind} (E,B) = \mu_{1}E^{2} + \mu_{2}B^{2} + \mu_{3}(\vec{E}\cdot\vec{B})\,,    
\label{constitutive relations 2}
\end{eqnarray}
where $\varepsilon_{1}$, $\varepsilon_{2}$, $\varepsilon_{3}$, $\mu_{1}$, $\mu_{2}$ 
and $\mu_{3}$ are related to the nonlinear medium 
dependence caused by the external fields presence. These quantities 
correspond to small disturbances when compared to the background 
values $\varepsilon_{0}$ and $\mu_{0}$ and they shall be considered, here and so on, only 
in the first order.  

In the classical hyperbolic partial differential equations theory, 
it is possible to analyze the propagation 
phenomenon by using the well known Hadamard-Papapetrou method 
\cite{ref14}. 
In such method, the propagation is considered by studying the evolution of the characteristic surfaces 
where the field is continuum. However, in general, the derivatives of the fields are not continuous. 
These surfaces can be understood as discontinuity surfaces  
which split the space-time into two globally separated domains. 
Let $f^{(1)}$ and $f^{(2)}$ be 
the two values of any function $f$ taken in the two domains. The Hadamard discontinuity 
of the function $f$ across a surface $\Sigma$ is defined as 
 $[f(x)]_{\Sigma} :=\displaystyle \lim_{\delta\to 0^{+}} \left\{f^{(1)}(x+\delta) 
- f^{(2)}(x-\delta)\right\}.$  In electrodynamics, for instance, the electromagnetic 
frontwave $\Sigma$ is supposed to be a surface through which the discontinuities of the electromagnetic field 
 (photons) propagate.  

Taking into account the electromagnetic fields as smooth functions across $\Sigma$ and their 
first derivatives with non null discontinuities, it is possible to apply the procedure described by 
the Hadammard-Papapetrou technique and get the 
following discontinuity conditions for each point $P\in\Sigma$ \cite{ref13}
\begin{eqnarray} \nonumber
[E^{\alpha}]_{\Sigma} = 0 ,\;\;\; [\partial_{\beta}E^{\alpha}]_{\Sigma} 
= e^{\alpha}k_{\beta} ,\;\;\; 
[\partial_{\beta}D^{\alpha}]_{\Sigma} =  
\left( \varepsilon e^{\alpha} - \acute{\varepsilon}\, E^{\alpha}E_{\gamma}e^{\gamma} 
- \dot{\varepsilon}\, E^{\alpha}B_{\gamma}b^{\gamma} \right) k_{\beta} ,
\end{eqnarray}
\begin{eqnarray}
[B^{\alpha}]_{\Sigma} = 0 ,\;\;\; [\partial_{\beta}B^{\alpha}]_{\Sigma} 
= b^{\alpha}k_{\beta} ,\;\;\; 
[\partial_{\beta}H^{\alpha}]_{\Sigma} = 
\Big( \dfrac{1}{\mu}\,b^{\alpha} + \dfrac{\acute{\mu}}{\mu^{2}}\, B^{\alpha}E_{\gamma}e^{\gamma} 
+ \dfrac{\dot{\mu}}{\mu^{2}}\, B^{\alpha} B_{\gamma}b^{\gamma} \Big) k_{\beta} \,,
\label{discontinuity EB2}
\end{eqnarray}
where we define $\acute{f}:=(1/E)({\partial} f/{\partial E})$ 
and $\dot{f}:=(1/B)({\partial} f/{\partial B})$. In the expressions 
presented above, the polarization vectors $e^{\alpha}$ and $b^{\alpha}$ 
are defined as $e^{\alpha}:=[\partial E^{\alpha}/\partial 
\Phi]_{\Sigma}$ and $b^{\alpha}:=[\partial B^{\alpha}/\partial \Phi]_{\Sigma}$. 
Moreover, $k^{\alpha}:=\partial^{\alpha}\Sigma=(\,\omega,\vec{q}\,)$ 
corresponds to the wave 4-vector orthogonal to $\Sigma$. 
The angular frequency of the electromagnetic waves and the 3-dimensional 
wave vector are given respectively by $\omega$ and $\vec{q}$.  

Applying the boundary conditions (\ref{discontinuity EB2}) in the 
equations (\ref{Maxwell Eqs.}), one can obtain the eigenvalues equation 
$Z^{\alpha}{}_{\beta} e^{\beta} = 0$, where the matrix $Z^{\alpha}{}_{\beta}$ 
can be written as \cite{ref13}
\begin{eqnarray}
Z^{\alpha}{}_{\beta} &=& \nonumber
\varepsilon h^{\alpha}{}_{\beta} - \acute{\varepsilon}{E}^{\alpha}{E}_{\beta}  
+ v^{-1}\Big[\dot{\varepsilon}(\,\hat{q}\times\vec{B}\,)_{\beta}{E}^{\alpha} 
+ \dfrac{\acute{\mu}}{\mu^{2}}(\,\hat{q}\times\vec{B}\,)^{\alpha}{E}_{\beta}\Big]
- \dfrac{v^{-2}}{\mu}\,(h^{\alpha}{}_{\beta}+\hat{q}^{\alpha}\hat{q}_{\beta}) 
\\&+& \dfrac{v^{-2}\dot{\mu}}{\mu^{2}}\Big[ 
(B^{2}- (\,\hat{q}\cdot\vec{B}\,)^{2})h^{\alpha}{}_{\beta} 
+ B^{2}\hat{q}^{\alpha}\hat{q}_{\beta} 
+ {B}^{\alpha}{B}_{\beta} - (\,\hat{q}\cdot\vec{B}\,)\hat{q}^{\alpha}{B}_{\beta}
- (\,\hat{q}\cdot\vec{B}\,)\hat{q}_{\beta}{B}^{\alpha}\Big] ,
\label{matrix Z}
\end{eqnarray}
where $v:=\omega/q$ is the phase velocity of the electromagnetic waves and 
$q^{\alpha}:=h^{\alpha}{}_{\beta}k^{\beta}=(0,\vec{q}\,)$. 
Nontrivial solutions ($e^{\alpha}\neq 0$) can be obtained if, and only if, 
$\det \mid Z^{\alpha}{}_{\beta} \mid = 0$, \cite{ref16}. The determinant is 
calculated only through the 3-dimensional structure $Z^{\alpha}{}_{\beta}$. 
The reason is that the time components of such matrix are identically null. 
This result is known as Fresnel generalized equation \cite{ref2, ref4} 
and it gives effective dispersion relations. These relations are a very 
useful tool in order to describe the light rays propagation inside 
the material media. 
\section{Birefringence: some new effects}
\label{secbir}
\hspace{0.65cm}The standard method to solve the Fresnel eigenvalues equation takes into account 
the eigenvector $e^{\alpha}$ expansion in a convenient base in the 3-dimensional 
space \cite{ref17}. A first aspect that arises here, despite the simplicity 
of this technique, is whether the set of chosen vectors is really linearly independent. 
In order to avoid this 
subtle aspect we shall apply here another method \cite{ref18} which deal only with  
the algebraic structure of the matrix $Z^{\alpha}{}_{\beta}$. In this approach, 
the 3-dimensional determinant of this matrix can be written in a covariant way 
\begin{equation}
\det \mid Z^{\alpha}{}_{\beta} \mid = -\frac{1}{6}(Z^{\alpha}{}_{\alpha})^{3} 
+\frac{1}{2} Z^{\alpha}{}_{\alpha}Z^{\beta}{}_{\gamma}Z^{\gamma}{}_{\beta} 
-\frac{1}{3} Z^{\alpha}{}_{\beta}Z^{\beta}{}_{\gamma}Z^{\gamma}{}_{\alpha}\,.
\end{equation}
The Fresnel generalized equation for the model studied here is given by 
the product of two polynomial as follows
\begin{eqnarray}
\Big[v^{2} \varepsilon - \dfrac{1}{\mu}\Big]\cdot\Big[v^{2} 
(\varepsilon 
+ \acute{\varepsilon}E^{2})  
+ v \Big(\dot{\varepsilon} + \dfrac{\acute{\mu}}{\mu^{2}} \Big) 
EB \sin \theta \sin \vartheta 
- \dfrac{\acute{\varepsilon}}{\varepsilon \mu}E^{2}\cos^{2}\phi 
+ \dfrac{\dot{\mu}}{\mu^2}B^{2}\sin^{2}\varphi - \dfrac{1}{\mu}\Big] 
= 0 ,
\label{dispersion relations}
\end{eqnarray}
where 
$\hat{q}\cdot\hat{E}=\cos\phi$, $\hat{q}\cdot\hat{B}=\cos\varphi$, $\hat{E}\cdot\hat{B}=\cos\theta$ 
and $(\hat{E}\times\hat{B})\cdot\hat{q}=\sin\theta\sin\vartheta$. We detach that $\vartheta$ 
is the angle formed by the vector $\hat{q}$ and the plane formed 
by the vectors $\hat{E}$ and $\hat{B}$. 

One can notice from the last equation that there are, 
in general, two dispersion relations for the possible polarization modes  
which propagate in the medium. This phenomenon is 
known as birefringence or double refraction. In what follows, we denote the polarization modes mentioned 
above as ordinary and extraordinary rays. Hence, the first and the second polynomial 
in Eq. (\ref{dispersion relations}) correspond to the dispersion relations  
for the ordinary and extraordinary rays, respectively. 
We derive in the Appendix, using the dispersion relations previously commented, the 
effective metric structures as well as the polarization modes for each ray.

As one can check, the phase velocity 
for the ordinary ray is written as  
\begin{eqnarray} 
v_{o}^{\pm} = \pm \frac{1}{\sqrt{\varepsilon_{0} \mu_{0}}}\Big[1 
-\dfrac{1}{2}\Big(\dfrac{\varepsilon_{1}}{\varepsilon_{0}}+\dfrac{\mu_{1}}{\mu_{0}}\Big)E^{2} 
- \dfrac{1}{2}\Big(\dfrac{\varepsilon_{2}}{\varepsilon_{0}}+\dfrac{\mu_{2}}{\mu_{0}}\Big)B^{2} 
-\dfrac{1}{2}\Big(\dfrac{\varepsilon_{3}}{\varepsilon_{0}}
+\dfrac{\mu_{3}}{\mu_{0}}\Big)EB\cos\theta\Big] \,.
\label{vo}
\end{eqnarray}

The quantity $v_{o}^{\pm}$ in last equation is isotropic, because it does not 
depend on the direction of propagation 
defined by the vector $\hat{q}$. On the other hand, the 
phase velocity of the extraordinary ray is composed by an isotropic part $v_{o}^{\pm}$ and 
an anisotropic one $\delta v^{\pm}$. The phase velocity of the extraordinary ray $v_{e}^{\pm}$ 
depends on the direction of propagation in the following way 
\begin{eqnarray}
v_{e}^{\pm} = v_{o}^{\pm} + \delta v^{\pm} , 
\label{ve}
\end{eqnarray}
where
\begin{eqnarray}\nonumber
\delta v^{\pm}&=&\Big[-\dfrac{\varepsilon_{3}\sin2\theta\sin\vartheta}{4\varepsilon_{0}} 
\pm \dfrac{\varepsilon_{1}\sin^{2}\phi}{\varepsilon_{0}\sqrt{\varepsilon_{0} 
\mu_{0}}}\Big] E^{2}+\Big[-\dfrac{\mu_{3}\sin2\theta\sin\vartheta}
{4\varepsilon_{0}{\mu_{0}}^{2}} \pm \frac{\mu_{2}\sin^{2}\varphi}
{\mu_{0}\sqrt{\varepsilon_{0} \mu_{0}}}\Big] B^{2}\\&+&\Big[
-\Big(\dfrac{\varepsilon_{2}}{\varepsilon_{0}} 
+ \dfrac{\mu_{1}}{\varepsilon_{0}{\mu_{0}}^{2}}\Big) \sin\theta\sin\vartheta 
\pm \frac{\cos\theta}{2\sqrt{\varepsilon_{0} \mu_{0}}}
\Big(\dfrac{\varepsilon_{3}}{\varepsilon_{0}}\sin^{2}\phi 
+ \dfrac{\mu_{3}}{\mu_{0}}\sin^{2}\varphi \Big)\Big] EB . 
\label{deltav}
\end{eqnarray}

Note that there are three distinct solution for the phase velocities. 
One of them corresponds to the ordinary ray. The another ones, which are not  
symmetric with respect to the direction of propagation, correspond 
to the extraordinary ray and are expected to represent 
two different velocities in the same direction. 

According to our previous discussion, 
$\varepsilon_{i}$ and $\mu_{i}$ represent  
small disturbances when compared to $\varepsilon_{0}$ and $\mu_{0}$. Hence,  
a straightforward comparison between the equations 
(\ref{vo}) and (\ref{deltav}) shows that $\mid v_{o} \mid >\mid \delta v \mid$.  
Consequently $v_{e}^{+}>0$ and $v_{e}^{-}<0$. This means  
that the phase velocity presented by the extraordinary ray,  
when it propagates in the direction defined by the vector $(\hat{q})$, is 
different from that one related to the propagation in the 
opposite direction defined by $(-\hat{q})$. Besides, 
since the ordinary ray propagates isotropically, 
birefringence phenomena will take place in any possible direction, 
except in the case where the phase velocity of both rays coincide \cite{ref13}. 

Let us remind the usual refraction index definition $(n:=v^{-1})$. Let us also 
define $\delta n^{\pm}:=( n_{o} - n_{e} )^{\pm}$, where $n_{o}$ and $n_{e}$ correspond 
to the refraction indexes of the medium related 
to the ordinary and extraordinary rays, respectively. 
From this point  
one can get $\delta n^{\pm} \cong \varepsilon_{0}\mu_{0}\delta v^{\pm}$.  
Using these definitions 
and Eq. (\ref{deltav}) one derive the following equation
\begin{eqnarray}\nonumber
\delta n^{\pm}&\cong&-\dfrac{\varepsilon_{3}\mu_{0}\sin2\theta\sin\vartheta}{4}\,E^{2} 
-\dfrac{\mu_{3}\sin2\theta\sin\vartheta}{4\mu_{0}}\,B^{2}
\pm \frac{\mu_3 \sqrt{\varepsilon_0}\cos\theta\sin^{2}\varphi}{2\sqrt{\mu_0}}\,EB\\
&\pm& \dfrac{\varepsilon_{1}\sqrt{\mu_{0}}\sin^{2}\phi}{\sqrt{\varepsilon_{0}}}\,E^{2}
\pm \dfrac{\mu_{2}\sqrt{\varepsilon_{0}}\sin^{2}\varphi}{\sqrt{\mu_{0}}}\,B^{2}
-\varepsilon_{2}\mu_{0}\sin\theta\sin\vartheta \,EB\nonumber\\
&+&\dfrac{\mu_{1}}{\mu_{0}} \sin\theta\sin\vartheta \,EB 
\pm\frac{\varepsilon_3\sqrt{\mu_0}\cos\theta\sin^{2}\phi }{2\sqrt{\varepsilon_0}}\,EB\,.
\label{deltan2}
\end{eqnarray}

The magnitude of the birefringence phenomena is expressed by this last equation 
for nonlinear dielectric isotropic media. The last five terms in Eq. (\ref{deltan2}) represent 
the known birefringence effects: Kerr \cite{ref7}, Cotton-Mouton \cite{ref8}, 
linear magnetoelectric \cite{ref19} and Jones \cite{ref9}. However,  
three new optical effects are represented by the three first terms. They  
emerge only due to the fact that we deal with the dielectric coefficients 
in the form (\ref{constitutive relations 1}) and (\ref{constitutive relations 2}). 

Let us discuss about the three first terms in the {\it r.h.s.} of 
Eq. (\ref{deltan2}). The first one is an electric-optical 
contribution\footnote{Electric-optical and 
magneto-optical birefringence occurs when the birefringence depends 
on the magnitude of the applied external electric and magnetic fields, respectively.}. 
One can see that an analogy between this term and the Kerr effect is straightforward, 
once the squared dependence on the electric field is observed in both of them. 
The second and the third ones are magneto-optical and magnetoelectric 
contributions, respectively. They can be understood as the analogue terms of the 
Cotton-Mouton and Jones effects, respectively. 
However, there is a difference between these three terms mentioned above 
and their analogue ones. If one consider the first term in {\it r.h.s.} 
for example, it is possible to observe that birefringence 
effect vanishes when the propagation is parallel to the plane formed by the 
electromagnetic fields and also when such fields are parallel\footnote{The Kerr 
analogue term depends on the direction of the magnetic field, but it does not depends on its magnitude.}. 
On the other hand, in the Kerr effect case, 
the direction of the magnetic field does not affect the measure of the 
electric-optical birefringence. The same observation can be 
extended for analogue cases of Cotton-Mouton and Jones effects. 
Furthermore, the birefringence magnitude will be maximizes when $\theta = \pi/4$  
(Kerr and Cotton-Mouton analogue terms) and $\theta =0$ (Jones analogue term). 

For the sake of completeness let us comment that the other five terms 
of Eq. (\ref{deltan2}) are nothing else but the well known 
birefringence effects already commented above. For example,  
the fourth and fifth terms in the {\it r.h.s.} 
correspond to Kerr and Cotton-Mouton effects, respectively. Observe that 
the intensity of such effects will be maximizes when the propagation is perpendicular  
to the direction defined by the applied field and it will be null if propagation is 
parallel to the electric and magnetic fields. In addition, 
the sixth and seventh terms in the {\it r.h.s.} correspond to the linear magnetoelectric 
birefringence. They achieve their maximum value in the crossed external 
fields situation ($\theta=\pi/2$) and their minimum (birefringence disappear) 
with parallel external fields ($\theta=0$). Moreover, the last term correspond to 
Jones effect with maximum value for crossed fields ($\theta=0$) and null for ($\theta=\pi/2$).

It should be remarked that magnetoelectric birefringence is 
composed by a pair of terms. The first pair (proportional to $\sin\theta$) represents 
magnetoelectric terms. The 
second pair (proportional to $\cos\theta$) is composed by the Jones effect and its 
analogue term (new one). Hence, as far as the angle $\theta$ between the  
external electromagnetic fields varies ($0\leq\theta\leq 2\pi$), one of 
these pairs decreases while the another one increases. Therefore, it is possible 
to infer from the previous analysis that 
the magnetoelectric birefringence will always take place.      
\section{Experimental tests, a perspective}
\label{testexp}
\hspace{0.65cm}
We present in this section a brief perspective about 
experimental tests, taking into account the whole birefringence 
scenario, including the three new terms presented in the Sec. 
\ref{secbir}. Some aspects of a consistent experimental  
framework about birefringence tests can be found in 
\cite{ref11}. Such a framework 
incorporates a successful experimental description of 
birefringence phenomenon. However, this subject is very much 
extensive and a deep approach here would not be compatible with the 
context of this work. 

In order to keep a better 
correspondence with the experimental scope, we begin by 
rewriting the 
quantity $\delta n$, described by the relation (\ref{deltan2}), in 
terms of the birefringence magnitude factor ${\cal K}(\la)$. For 
example, the $\delta n$ contribution for the Kerr, Cotton-Mouton,  
Jones and linear magnetoelectric effects may be written as follows 
\begin{equation}
\delta n_{K}={\cal K}_{K}(\la)\,\la\,\sin^2\phi\,E^2\,,\hspace{1.0cm} 
\delta n_{CM}={\cal K}_{CM}(\la)\,\la\,\sin^2\varphi\, B^2\,, \label{kcm}
\end{equation}
\begin{equation}
\delta n_{J}={\cal K}_{J}(\la)\, \la\,\cos{\theta}\sin^2\phi \,EB\,,
\hspace{1.0cm}
\delta n_{ME}= {\cal K}_{ME}(\la)\, \la\, \sin\theta\,\sin\vartheta EB\,,
\label{mejones}
\end{equation}
where $\la$ is the wavelength of the light propagating in the medium. 
A comparison between relations (\ref{deltan2}), (\ref{kcm}) and (\ref{mejones}) 
allows one to write the birefringence magnitude factors presented in the last equations, in the form   
\begin{equation}
{\cal K}_{K}=\frac{\epsilon_1\sqrt{\mu_0}}{\la\,\sqrt{\epsilon_0}}\,,
\hspace{0.8cm}
{\cal K}_{CM}=\frac{\mu_2\sqrt{\epsilon_0}}{\la\sqrt{\mu_0}}\,,
\hspace{0.8cm}
{\cal K}_{J}=\frac{\epsilon_3 \sqrt{\mu_0}}{2\la\sqrt{\epsilon_0}}\,,
\hspace{0.8cm}
{\cal K}_{ME}=\frac{\mu_0\epsilon_{2}}{\la}\,.
\end{equation}
All these effects represented in (\ref{kcm}) and (\ref{mejones}) 
has already been experimentally confirmed \cite{ref11}. 
However, the search for 
material media where birefringence phenomena are detectable with good 
accuracy is not quite trivial. Usually, for a particular effect,  
a set of measurements are performed for different materials. 
The effects are better observed  in the medium which presents 
a grater value  for the quantity ${\cal K}(\la)$. 

Concerning to Jones and magnetoelectric birefringence for example, 
the measurements were performed  for a set of materials and a table relating 
these materials and their corresponding values for the quantities 
${\cal K}_{J}$ and ${\cal K}_{ME}$ can be found in 
\cite{ref11}. A sample 
of methylcyclopentadienyl-Mn-tricarbonyl presents the most relevant 
results: ${\cal K}_{J}= 47\times 10^{-12}\, \mbox{V}^{-1}\,\mbox{T}^{-1}$ 
and ${\cal K}_{ME}=51\times 10^{-12}\, \mbox{V}^{-1}\,\mbox{T}^{-1}$. 
As the authors have detached in \cite{ref11}, 
such values confirm the validity of the equations (\ref{mejones}),  
proving the existence of Jones and  magnetoelectric birefringence.  
Furthermore, results for the Kerr and Cotton-Mouton effects agree reasonably 
with the literature values and the validity of the equations (\ref{kcm}) has also been 
confirmed. In particular, a sample of nitrobenzene shows the following 
acceptable values ${\cal K}_{K}=3,9\times 10^{-12}\,\mbox{m}\,\mbox{V}^{-2}$ and 
${\cal K}_{CM}=2,1\times 10^{-2}\,\mbox{m}^{-1}\,\mbox{T}^{-2}$. 

As pointed out in Sec. (\ref{secbir}), the three new theoretical effects 
predicted by the approach considered in this paper represent the analogue terms of 
Kerr, Cotton-Mouton and Jones effects. We shall denote their corresponding 
birefringence magnitude factors as ${\cal K}_{KA}$, ${\cal K}_{CMA}$ and 
${\cal K}_{JA}$, respectively. In the same way as performed above, one can write 
these quantities as follows   
\begin{equation}
{\cal K}_{KA}=\frac{\varepsilon_2\mu_0}{4\la}\,,\hspace{0.8cm}
{\cal K}_{CMA}=\frac{\mu_3}{4\la\mu_0}\,,\hspace{0.8cm}
{\cal K}_{JA}=\frac{\mu_3\sqrt{\epsilon_0}}{2\la\sqrt{\mu_0}}\,.
\end{equation}

Concerning the experimental observations of these effects, one 
has to take into account that the well understanding about Jones 
birefringence and molecular structure is not complete yet. Hence 
pure molecular liquids are good candidates to perform measurements 
of the quantity ${\cal K}_{JA}$ presented in the last equation. 
Notwithstanding, methylcyclopentadienyl-Mn-tricarbonyl 
seems to be a remarkable sample when the birefringence 
effect (to be measured) is proportional to $EB$. 
For these reasons we believe that it presents good 
perspectives in order to confirm Jones analogue effect. 
By similar reasons, samples of nitrobenzene seem to 
play the role as good candidates in order to perform measurements of the 
Kerr and Cotton-Mouton analogue effects.

As a last comment we detach that very technical issues 
are related to the descriptions and details 
about the experimental setup to observe the new effects 
described here and, as we already stressed, 
considerations about the merit of these questions 
would not be in the context of this work.
\section{Conclusions and discussions}
\label{Con}
\hspace{0.65cm}
The propagation description of the electromagnetic monochromatic waves 
was considered in the context of isotropic nonlinear dielectric 
media in the eikonal limit of the geometric optics. By using the 
Hadamard-Papapetrou technique, the eigenvalues problem 
was presented and solved for isotropic material media represented by the 
dielectric coefficients  $\varepsilon=\varepsilon(E,B)$ and $\mu=\mu(E,B)$. 
We found the dispersion relations and the effective optical 
metric structure related to each possible polarization mode. 

It is worth noting that the birefringence study we performed, taking into 
account the dielectric coefficients mentioned above, produced a new result which 
is represented in Eq. (\ref{deltan2}). According to this equation, there are three 
new optical effects unknown until now. One of them presents a squared dependence 
on electric field and for this reason it can be understood as an 
analogue term of the Kerr effect. By the same way, the another two terms 
can be interpreted as the analogues ones of 
the Cotton-Mouton and Jones effects. Furthermore, Eq. (\ref{deltan2}) also 
contemplates the five well known effects presented in the literature: 
Kerr, Cotton-Mouton, Jones and linear birefringence magnetic-electric. 

We have also addressed in Sec. (\ref{testexp}) the possibility of performing experimental tests 
in order to confirm the three new theoretical effects presented in section 
Sec. (\ref{secbir}). The scenario for pure molecular liquids seems to be promising.  
In particular, samples of methylcyclopentadienyl-Mn-tricarbonyl are good candidates  
in order to measure the Jones analogue effect. Nitrobenzene seems to be 
good samples in the measurements of the Kerr and Cotton-Mouton analogue effects.  
\section*{Acknowledgments}
\hspace{0.65cm}This work was partially supported by the 
Brazilian research agencies  
Conselho Nacional de Desenvolvimento Cient\' ifico e Tecnol\' ogico (CNPq),
Funda\c c\~ ao de Amparo  a Pesquisa do Estado de Minas Gerais (FAPEMIG) and  
Funda\c c\~ ao Nacional de Desenvolvimento da Educa\c c\~ ao (FNDE). 
DP and BR are grateful to CEFET-RJ and CEFET-MG, repectively,
by the structure and support offered to perform the work. 
The authors dedicate the work to Guilherme P. Goulart (in memory). 
\section*{Appendix}
\hspace{0.65cm}

\hspace{0.65cm}We present in this Appendix two additional aspects related 
with the model treated here:  
the effective metric for the ordinary and  
extraordinary rays, as well as a brief description of the polarization modes 
in the context of the Kerr effect.
\vspace{0.3cm}

{\large {\bf Effective geometry}}\\ 
\hspace{0.65cm} Let us begin mentioning that the development of methods which enable one to test several 
cinematic aspects of General Relativity  in the laboratory has been 
explored in some areas of physics. For example, in nonlinear electrodynamics 
it has been considered as a possible scenario to construct analogue models 
of General Relativity, in the context of nonlinear Lagrangian and 
nonlinear material media. It is based on the fact that the 
trajectory of photons can be described in terms of null geodesics in a effective geometry  
 ${\cal{G}}^{\alpha\beta}$, also known as optical geometry  
(see \cite{ref15, ref20} for more information about 
effective geometry approach in the context of nonlinear electrodynamics).

For several physical configurations, the dispersion relations can be written in a 
suggestive way as 
${\cal{G}}^{\alpha\beta}k_{\alpha}k_{\beta}=0$. In the model considered here, we 
write the dispersion relation for the ordinary ray in the form
\begin{eqnarray} 
\omega^{2} \varepsilon\mu - q^{2} = 0 = ({\cal{G}}_{o}){}^{\alpha\beta}k_{\alpha}k_{\beta} ,
\label{dr}
\end{eqnarray}
where
\begin{eqnarray} 
({\cal{G}}_{o}){}^{\alpha\beta} = \eta^{\alpha\beta} - (1 - \varepsilon\mu)V^{\alpha}V^{\beta} ,
\label{geometry effective o}
\end{eqnarray}
represents the optical metric related to the ordinary ray and corresponds to the 
Gordon's metric \cite{ref21}. In the same way, for the extraordinary ray we get 
\begin{eqnarray}
({\cal{G}}_{e}){}^{\alpha\beta} = ({\cal{G}}_{o}){}^{\alpha\beta} +\Big(\acute{\varepsilon}\mu E^{2}
+\varepsilon\dot{\mu}B^{2}\Big)V^{\alpha}V^{\beta} - \dfrac{\acute{\varepsilon}}{\varepsilon}
E^{\alpha}E^{\beta} - \dfrac{\dot{\mu}}{\mu}B^{\alpha}B^{\beta} + 
\dfrac{1}{2}\Big(\dot{\varepsilon}\mu+\dfrac{\acute{\mu}}{\mu}\Big)
(\vec{E}\times\vec{B})^{(\alpha}V^{\beta)} .
\label{geometry effective e}
\end{eqnarray}
Observe that such an effective geometries indeed correspond to the 
background Minkowskian metric deviations. Hence, the nonlinear properties of the 
medium do affect the trajectories of the light ray.

\vspace{0.4cm}
{\large {\bf Polarization: Kerr effect}\\ 
\hspace{0.65cm}For any one of the 
cases cited in the Sec. (\ref{secbir}), one can get the corresponding polarization vector 
($e^{\alpha}$, $b^{\alpha}$) by leading any specific solution in the 
generalized Fresnel equation. We consider the following electromagnetic 
configuration \{$\varepsilon=\varepsilon(E)$; $\mu=\mu_{0}$; $\hat{E}
=\hat{x}$; $\hat{B}=\hat{y}$; $\hat{q}=\hat{z}$\}, wich is the source 
of the Kerr effect with crossed external electromagnetic 
fields and a perpendicular propagation to the plane formed by such fields. 
Hence, by writing the polarization vector in the 3-dimensional representation  
$\hat{e} = c_{1} \hat{x} + c_{2} \hat{y} + c_{3} \hat{z}$ and considering the 
generalized Fresnel equation, together the dispersion relation for each ray, 
one can find the following polarization vectors  
\begin{eqnarray}
(\hat{e}_{o},\hat{b}_{o}) = (\hat{y},-\hat{x}) = (\hat{B},-\hat{E}) , 
\hspace{0.8cm} (\hat{e}_{e},\hat{b}_{e}) = (\hat{x},\hat{y}) = (\hat{E},\hat{B}) ,
\label{polarization 1}
\end{eqnarray}
where we got the vestors $\vec{b}$ by using the relation  
$\vec{b}=\omega^{-1}(\,\vec{q}\times\vec{e}\,)$. The result presented above can 
also be written in the matrix form
\begin{eqnarray}
\left(
\begin{array}{c}
\hat{e}_{o} \\ \hat{b}_{o}
\end{array}
\right) = 
\left(
\begin{array}{cc}
0 & 1 \\
-1 & 0 \\
\end{array}
\right)
\left(
\begin{array}{c}
\hat{e}_{e} \\ \hat{b}_{e}
\end{array}
\right) .
\label{polarization 2}
\end{eqnarray}

Observe that in the Kerr effect, for example, one can get the corresponding polarization vectors for the 
ordinary ray in terms of the  polarization vectors related to the 
extraordinary ray by using a rotation in the plane formed by the external electromagnetic fields. 
The extension of the procedure considered here to any electromagnetic configuration presented in the 
Sec. (\ref{secbir}) is straightforward.  
\renewcommand{\baselinestretch}{0.9}

\end{document}